\documentclass[10pt,twocolumn,preprintnumbers,amsmath,amssymb,nofootinbib
,superscriptaddress]{revtex4-2}
\usepackage{amsmath,amssymb,amsfonts}
\usepackage{hyperref}
\usepackage{graphicx}
\usepackage{subfig}
\usepackage{lipsum}
\usepackage{diagbox} 
\usepackage{array}
\usepackage{float}
\usepackage{xcolor}
\usepackage[normalem]{ulem}

\begin{document}

\title{Quantum Corrections to Randall-Sundrum Model from JT Gravity}

\author{Ying-Jian Chen}
\email{chenyingjian22@mails.ucas.ac.cn}
\affiliation{International Centre for Theoretical Physics Asia-Pacific (ICTP-AP), University of Chinese Academy of Sciences (UCAS), Beijing 100190, China}
\affiliation{Taiji Laboratory for Gravitational Wave Universe (Beijing/Hangzhou), UCAS, Beijing 100049, China}
    \affiliation{Center for Gravitational Wave Experiment, National Microgravity Laboratory, Institute of Mechanics, Chinese Academy of Sciences, Beijing 100190, China}

\author{Jun Nian}
\affiliation{International Centre for Theoretical Physics Asia-Pacific (ICTP-AP), University of Chinese Academy of Sciences (UCAS), Beijing 100190, China}

\begin{abstract}
\noindent
We investigate quantum corrections to the Randall-Sundrum (RS) model in the near-extremal black brane background with quantum corrections in the near-horizon. The near-horizon geometry is described by Jackiw-Teitelboim gravity, and the quantum fluctuations are governed by the Schwarzian action. We introduce the Schwarzian modes into the RS metric, derive the quantum-corrected equation for the Kaluza-Klein (KK) modes via the Schwinger-Dyson equation, calculate the correction to the KK mass spectrum, and discuss the impact of quantum corrections on the Goldberger-Wise mechanism. Our work introduces both quantum corrections and temperature into the RS model, providing insights into cosmology and phase transitions within it.
\end{abstract}

\maketitle

\section{Introduction}

The quest to unify the fundamental forces of nature and resolve the hierarchy problem, one of the most important problems in particle physics, has spurred the development of numerous higher-dimensional theoretical frameworks \cite{DOMAIN-WALL, EXTRA1, WOS:000074694500008}. Among these, the Randall-Sundrum (RS) model, proposed by Lisa Randall and Raman Sundrum in 1999 \cite{Large, alternative}, stands as the first theory of warped extra dimensions, proposes a potential solution to the hierarchy problem, and has attracted attention in the academic community.

At its core, the RS model assumes a five-dimensional anti-de Sitter (AdS$_5$) bulk spacetime bounded by two 3-branes: the ultraviolet (UV) brane and the infrared (IR) brane. A key feature of this framework is the warp factor, an exponential function of the extra-dimensional coordinate that naturally red-shifts the Planck scale from the UV brane to the IR brane, thereby addressing the hierarchy between the gravitational and electroweak scales without invoking unnaturally large compactification volumes and other hierarchies. Furthermore, this model also predicts the mass spectrum of Kaluza-Klein (KK) gravitons, indicating that the masses of their excited states are on the TeV scale, and their coupling strengths are suppressed to the TeV scale. Therefore, this model is subject to experimental verification. After the proposal of the RS model, there have been numerous developments and applications.

The RS model exhibits a deep connection with the AdS/CFT correspondence \cite{WOS:000081063400007}, as the extra-dimensional evolution along the bulk maps directly to the renormalization group flow of a four-dimensional nearly conformal field theory (CFT) in the dual picture. A critical dynamical degree of freedom in the RS model is the radion, a massless mode associated with fluctuations of the radius of extra dimension, whose stabilization is described by the Goldberger-Wise (GW) mechanism \cite{WOS:000084152000004}. Recent renewed interest in the RS model stems from its implications for early universe cosmology, including the generation of a stochastic gravitational wave background via supercooled phase transitions \cite{WOS:000175277100051, WOS:001116716600002, WOS:000423800200001}. However, recent studies have indicated that it is a cooled phase transition rather than supercooled \cite{WOS:001251818200003}, further underscoring its relevance to both theoretical and observational physics.

However, the RS model still has some issues at present. The model's fundamental assumption is a smooth geometric structure, making it purely classical, and the influence of quantum gravity effects has not been considered so far. Furthermore, the model itself does not incorporate temperature, which hinders rigorous treatments of phase transitions within the RS model. It is worth noting that the potential solutions to these issues are associated with Jackiw-Teitelboim (JT) gravity and its linkage to Schwarzian modes.

 The JT gravity \cite{JACKIW, twoSPACETIME} is a two-dimensional gravitational theory characterized by a simple action involving an Einstein-Hilbert term and a constant negative cosmological constant in the bulk as well as a Gibbons-Hawking-York term on the boundary. It has emerged as a powerful tool for describing the near-horizon dynamics of higher-dimensional near-extremal Reissner-Nordstr\"om (RN) black holes \cite{Turiaci}. A defining insight is that the collective behavior of the near-horizon region of such a black hole, where the spacetime geometry reduces to AdS$_2$, is effectively captured by a JT-like gravitational action, with the boundary dynamics governed by a Schwarzian action \cite{Zhenbin}. The Schwarzian mode, arising from the spontaneous breaking of conformal symmetry in the AdS$_2$ boundary, encodes crucial information about the thermodynamics of the underlying black hole, including its statistical entropy and free energy, which are correctly reproduced by the partition function of the Schwarzian action. For the Schwarzian action, we can perform a path integral, allowing us to consider quantum fluctuations in the near-horizon region of near-extremal black holes. This kind of quantum fluctuation stands out because thermal fluctuations are suppressed; we refer to it as an infrared quantum fluctuation. Recently, there have been papers discussing the corrections to various physical quantities caused by this type of quantum fluctuations \cite{liu2024quantumcorrectionsholographicstrange, Brown:2024ajk, liu2025quantumcorrectedholographicwilsonloop, Emparan:2025sao, zheng, nian2025quantumcorrectionslowtemperaturefluidgravity, PandoZayas:2025snm, cremonini2025quantumcorrectionsetasjt}. In our paper, we investigate the quantum corrections to the RS model due to the quantum fluctuations of the near-extremal RN black brane.

 The plan of this paper is as follows. In Sec.~\ref{RS}, we review some basic ingredients of the RS model. In Sec.~\ref{JT gra}, we introduce the main tools of this paper, the JT gravity and Schwarzian modes, and calculate the quantum correction factor to the AdS$_2$ metric by incorporating Schwarzian modes. In Sec.~\ref{RSQ}, we discuss quantum corrections to the RS model by considering a near-extremal RN black brane and invoking Schwarzian modes in its near-horizon region. Using the quantum-corrected RS model, we reexamine the Kaluza-Klein mass spectrum and the GW mechanism with quantum corrections in Sec.~\ref{GM} and Sec.~\ref{GW}, respectively. Finally, the summary and outlook are presented in Sec.~\ref{con}.

\section{Randall-Sundrum model}\label{RS}

The Randall-Sundrum (RS) model posits a single extra dimension that is compactified on a circle, with the upper and lower halves of the circle identified to construct the $S^1/Z_2$ orbifold \cite{Large}. Denoted by the coordinate $y$, this extra dimension possesses two fixed points located at  $y=0$ and $y=\pi r_c$, where $r_c$ serves as the radius of the compactified extra dimension. Each fixed point is inhabited by a four-dimensional brane (referred to as a 3-brane), and these two 3-branes, spaced $\pi r_c$ apart, confine a five-dimensional (5D) bulk spacetime. The total action of the model is composed of three parts: one arising from gravity in the 5D bulk, and the other two encoding the physics localized on each of the two 3-branes, i.e.,
 \begin{equation}
     \begin{aligned}
         S_G & = \int d^4x\int_{-\pi r_c}^{\pi r_c}dy\sqrt{-G}(2M^3R-\Lambda)\, ,\\
         S_{UV} & = \int d^4x\sqrt{-g_0}(\mathcal{L}_{UV}-V_{UV})\, ,\\
         S_{IR} & = \int d^4x\sqrt{-g_{\pi r_c}}(\mathcal{L}_{IR}-V_{IR})\, ,
     \end{aligned}
 \end{equation}
where $g_0$ and $g_{\pi r_c}$ represent the determinants of the induced metrics of the 5D metric on the two branes, respectively. $\mathcal{L}$ represents the Lagrangian of the matter field on the brane, while $V$ is the energy density of the brane. The spacetime metric $G_{MN}$ is \cite{Large}
\begin{equation}
    ds^2_{RS} = e^{-2k|y|}\eta_{\mu\nu}dx^\mu dx^\nu+dy^2\, .
\end{equation}
Note that the spacetime in between the two 3-branes is simply a slice of an AdS$_5$ geometry. According to the AdS/CFT correspondence, the 5D bulk spacetime is dual to the 4D boundary conformal field theory \cite{WOS:000081063400007}. The two branes correspond to the ultraviolet and the infrared cutoffs of the field theory. We use subscripts UV and IR to label the actions of the two branes located at $y = 0$ and $y= \pi r_c$, respectively.

The metric $G_{MN}$ is conformally flat. To see it, we can make a change of coordinate 
\begin{equation}
    dy^2 = e^{-2k|y|} dz^2\, .
\end{equation}
Subsequently, the metric becomes
\begin{equation}
    ds^2_{RS}=\frac{1}{(k|z|+1)^2}\, \eta_{MN}\, dx^M\, dx^N\, .
\end{equation}
To understand gravitational dynamics in the RS model, we must derive explicit expressions for gravitons, particles associated with small perturbations $h_{MN}(x,y)$ around the background metric
\begin{equation}
    G_{MN} = \frac{1}{(k|z|+1)^2} (\eta_{MN}+h_{MN})\, .
\end{equation}
It is particularly convenient to work with a gauge in which the fluctuations do not have any extra-dimension component and are transverse and traceless, i.e.,
\begin{equation}
    \begin{aligned}
        h_{Mz} & = 0\, ,\\
        \partial_\mu h^{\mu\nu} & = 0\, ,\\
        h^\mu\,_\mu & = 0\, .
    \end{aligned}
\end{equation}
The linearized Einstein equation in $y$ coordinate is \cite{alternative}
\begin{equation}
    \partial_\rho\partial^\rho h_{\mu\nu} + e^{2ky}\partial_y(e^{-4ky}\partial^yh_{\mu\nu}) = 0\, .\label{Ein}
\end{equation}
Performing a separation of variables on this equation
\begin{equation}
    h_{\mu\nu}(x^\mu,y) = \Sigma_{n=0}^{n=\infty} h_{\mu\nu}^{n}(x^\mu) f_n(y)\, ,
\end{equation}
we obtain the following equations
\begin{equation}
    \begin{aligned}
        \partial_\rho\partial^\rho h^n_{\mu\nu} & = m_n^2h_{\mu\nu}^n\, ,\\
        -\partial_y(e^{-4ky}\partial^yf_n) & = m_n^2e^{-2ky}f_n\, ,
    \end{aligned}
\end{equation}
whose boundary conditions are
\begin{equation}
    \begin{aligned}
        \partial_yf_n|_{y=0} & = 0\, ,\\
       \partial_yf_n|_{y=\pi r_c} & = 0\, ,\label{b}
    \end{aligned}
\end{equation}
and $f_0=const$ for $m_0=0$. For $m_n\neq0$, the general solution of the KK mode can be written as \cite{alternative}
\begin{equation}
    f_n = N_ne^{2ky} \left[J_2\left(\frac{m_n}{k}e^{ky}\right) + b_nY_2\left(\frac{m_n}{k}e^{ky}\right)\right]\, ,
\end{equation}
where $N_n$ and $b_n$ are constants. Correspondingly, the KK mass spectrum is \cite{alternative}
\begin{equation}
    m_n\approx(n+\frac{1}{4})\pi k e^{-k\pi r_c}\, .\label{m}
\end{equation}
The RS model assumes that the fundamental mass scale is the Planck scale. To solve the hierarchy problem, the parameter $k$ with mass dimension should also be on the Planck scale, and the effective mass scale on the 4D IR brane must be redshifted down to the TeV scale relative to the 5D fundamental scale. Consequently, $k e^{-k\pi r_c}$ is $\mathcal{O} (1\, \mathrm{TeV})$, and the KK gravitons in the RS model have masses of the TeV scale.

Up to this point, the radius of the extra dimension has been treated as a free parameter. However, this degree of freedom implies a massless scalar field in the effective theory, corresponding to fluctuations of the radius along the extra dimension. This massless scalar field would induce a fifth force in violation of the equivalence principle. Therefore, to preserve the viability of the Randall-Sundrum model, the scalar field has to obtain a mass, i.e., to be stabilized. A well-established approach to achieve this is the Goldberger-Wise mechanism \cite{WOS:000084152000004}. Classically, a potential emerges from a bulk scalar field with interaction terms localized on the two 3-branes. This potential's minimum can be tuned to yield $k r_c\sim 10$ without recourse to fine-tuning of parameters \cite{WOS:000084152000004}.

\section{JT gravity and quantum correction to A\NoCaseChange{dS}$_2$}\label{JT gra}

Let us consider the Euclidean Jackiw-Teitelboim (JT) gravity action \cite{JACKIW, twoSPACETIME}:
\begin{align}
  I_{JT} & = -\frac{1}{16\pi G_N}\int d^2 x\, \sqrt{g}\phi(R_c+2) \nonumber\\
  {} & \quad - \frac{1}{8\pi G_N}\int_{bdy} \sqrt{g_b}\phi_b(K-1)\, ,
\end{align}
where $R_c$ is the scalar curvature, and $K$ is the boundary extrinsic curvature. We first vary the action $I_{JT}$ with respect to the dilaton field $\phi$, leading to $R_c=-2$, which implies that the bulk is a patch of the 2d hyperbolic surface $ds^2 = \frac{dT^2+dZ^2}{Z^2}$. Dynamics is now entirely dictated by the Gibbons-Hawking-York (GHY) boundary term on a boundary curve. This can be visualized on the hyperbolic upper half surface, where a region close to the actual boundary is cut out. We parametrize the boundary curve as $(T=F(s), Z(s))$, where $s$ is the proper time on the boundary. Fixing the boundary metric to $\frac{1}{\epsilon}$ imposes $Z(s)=\epsilon F'(s)$. We can evaluate the extrinsic curvature \cite{Zhenbin}:
\begin{equation}
    \begin{aligned}
        K & = 1 + \epsilon^2[F,s] + O(\epsilon^4)\, ,\\
        \text{with}\quad [F,s] & \equiv\frac{F'''}{F'} - \frac{3}{2}\left(\frac{F''}{F'}\right)^2\, .
    \end{aligned}
\end{equation}
Fixing the boundary dilaton field $\phi_b\equiv\frac{a}{2\epsilon}$ yields the Schwarzian action as follows: 
\begin{equation}
  \begin{aligned}
        I_{Schw}[F] & = -C\int ds[F,s]\, ,\\
        \text{with}\quad C & \equiv\frac{a}{16\pi G_N}\, .
  \end{aligned}
\end{equation}
The Schwarzian action treats the reparametrization $F(s)$ as the dynamical degree of freedom. It turns out that adopting an alternative reference frame is more natural when analyzing thermal systems. We define
\begin{equation}
   F(s)\equiv \tan\frac{\pi}{\beta}f(s)\, ,
\end{equation}
where $f(s)$ is a new variable with the properties: 
\begin{equation}
    f(s+\beta)=f(s)+\beta\, ,\quad f'(s)\geq 0\, .
\end{equation}
These two properties enable interpreting $f(s)$ as a time reparametrization of the boundary thermal circle. Here, $\beta$ is the inverse temperature, and $f(s)=s$ corresponds to the classical solution.

For the quantum case, we should consider all different reparameterizations when performing the path integral. However, the exact computation can be quite difficult. Instead, we can apply the perturbation theory by expanding the reparametrization $f(s)$ around its saddle 
\begin{equation}
    f(s)=s+\epsilon(s)\, ,\quad \text{with } \epsilon(s+\beta)=\epsilon(s)\, ,
\end{equation}
where $\epsilon(s)$ is referred to as the Schwarzian modes. Consider the AdS$_2$ metric in the light-cone coordinates 
\begin{equation}
    ds^2_{AdS_2}=-\frac{4\pi^2}{\beta^2}\frac{L^2}{12}\frac{dx^+dx^-}{\sin^2{\frac{\pi}{\beta}(x^+-x^-)}}\, ,\label{AdS2}
\end{equation}
where $\frac{L}{\sqrt{12}}$ is the radius of AdS$_2$. We can introduce the Schwarzian modes into this classical metric
\begin{equation}
    \begin{aligned}
        {} & ds^2_{AdS_2f} \\
        =\, & -\frac{4\pi^2}{\beta^2}\frac{L^2}{12}\frac{f'(x^+)f'(x^-)}{\sin^2\frac{\pi}{\beta}(f(x^+)-f(x^-))}dx^+dx^-\\
        =\, & -\frac{4\pi^2}{\beta^2}\frac{L^2}{12}\frac{(1+\epsilon'(x^+))(1+\epsilon'(x^-))}{\sin^2\frac{\pi}{\beta}((x^+-x^-)+(\epsilon(x^+)-\epsilon(x^-)))}dx^+dx^-\\
        \equiv\, & \frac{1}{A_{cor}} ds^2_{AdS_2}\, ,
    \end{aligned}
\end{equation}
where we have defined a quantum correction factor $A_{cor}$, which has the following expansion up to the second order of Schwarzian modes:
\vspace{5mm}
\begin{widetext}
    \begin{equation}
        \begin{aligned}
             A_{cor} =\, & 1+\frac{2\pi}{\beta}(\epsilon(x^+)-\epsilon(x^-))\cot\frac{\pi}{\beta}(x^+-x^-) -(\epsilon'(x^+)+\epsilon'(x^-))+(\epsilon'(x^+)+\epsilon'(x^-))^2-\epsilon'(x^+)\epsilon'(x^-) \\
        & - \frac{2\pi}{\beta}(\epsilon'(x^+)+\epsilon'(x^-))(\epsilon(x^+)-\epsilon(x^-))\cot\frac{\pi}{\beta}(x^+-x^-)+\frac{\pi^2}{\beta^2}\frac{\cos\frac{2\pi}{\beta}(x^+-x^-)}{\sin^2\frac{\pi}{\beta}(x^+-x^-)}(\epsilon(x^+)-\epsilon(x^-))^2\, .\label{A}
        \end{aligned}
    \end{equation}
\end{widetext}

\section{RS metric with quantum corrections}\label{RSQ}

We consider a near-extremal Reissner-Nordstr\"om (RN) black brane in a five-dimensional anti-de Sitter (AdS) spacetime:
\begin{equation}
    \begin{aligned}
        ds^2_{RN} & = -\frac{u^2}{L^2}f(u)dt^2 + \frac{L^2}{u^2f(u)}du^2 + \frac{u^2}{L^2}(dx_1^2+dx_2^2+dx_3^2)\, ,\\
    f(u) & = 1 - (1+Q^2)\frac{u_T^4}{u^4}+Q^2\frac{u_T^6}{u^6}\, ,\label{full}
    \end{aligned}
\end{equation}
where $L$ is the radius of the five-dimensional AdS spacetime, $u_T$ is the location of the horizon, and $Q$ is the charge of the black brane.

Taking the limit $u\gg u_T$, we obtain the metric in the region far from the black brane horizon:
\begin{equation}
     ds^2_{FAR} = -\frac{u^2}{L^2}dt^2 + \frac{L^2}{u^2}du^2 + \frac{u^2}{L^2}(dx_1^2+dx_2^2+dx_3^2)\, .\label{eq:FAR class metric}
\end{equation}
We can perform the coordinate transformation:
\begin{equation}
    \begin{aligned}
    y & = -L\ln\frac{u}{u_0}\, ,\\
    t' & = \frac{u_0}{L} t\, ,\\
    x' & = \frac{u_0}{L} x\, ,
    \end{aligned}
\end{equation}
where $u_0$ is a constant parameter. Removing the prime, we obtain a new metric:
\begin{equation}
    ds^2_{FAR} = dy^2+e^{-\frac{2y}{L}}(-dt^2+dx_1^2+dx_2^2+dx_3^2)\, .
\end{equation}
This is exactly the standard form of the RS metric. The parameter $u_0$ denotes the location of the UV brane.

Now, let us consider the near-horizon region of the black brane. By taking the limit $u-u_T\ll u_T$, we obtain the metric in the near-horizon region: 
    \begin{equation}
   \begin{aligned}
        ds^2_{NHR} & = -\frac{12(R^2-\delta^2)}{L^2}dt^2 + \frac{L^2}{12(R^2-\delta^2)}dR^2\\
        {} & \quad + \frac{u_T^2}{L^2}(dx_1^2+dx_2^2+dx_3^2)\, ,\\
        \text{with}\quad R & \equiv u-u_T+\delta\, ,\quad \delta \equiv \frac{L^2\pi}{6\beta}\, ,\\
        \beta & \equiv \frac{2L^2\pi}{(2-Q^2)u_T}\, ,
   \end{aligned}
\end{equation}
where $\beta$ also denotes the inverse temperature of the black brane. Through the coordinate transformation $R=\delta \coth\frac{2\pi}{\beta}w$, the metric can be expressed as 
\begin{equation}
    ds^2_{NHR}=\frac{4\pi^2}{\beta^2}\frac{L^2}{12}\frac{dw^2-dt^2}{\sinh^2\frac{2\pi}{\beta}w}+\frac{u_T^2}{L^2}(dx_1^2+dx_2^2+dx_3^2)\, .
\end{equation}
We see that the near-horizon geometry of the 5D near-extremal black brane is a direct product of AdS$_2$ and $\mathbb{R}^3$ or T$^3$, if we impose the periodic boundary conditions on the boundary spatial dimensions. Moreover, we can read off the radius of AdS$_2$ as $\frac{L}{\sqrt{12}}$. If we perform a Wick rotation $\tau = it$ and introduce the light-cone coordinates
\begin{equation}
    \begin{aligned}
        x^+ & \equiv iw-\tau\, ,\\
        x^- & \equiv -iw-\tau\, ,
    \end{aligned}
\end{equation}
the AdS$_2$ part of the near-horizon metric can be transformed into \eqref{AdS2}. Therefore, after turning on the Schwarzian modes in the AdS$_2$ part, we obtain a new near-horizon metric with quantum fluctuations:
\begin{align}
    ds^2_{NHRf} & = -\frac{12(R^2-\delta^2)}{L^2}\frac{1}{A_{cor}}dt^2 + \frac{L^2}{12(R^2-\delta^2)}\frac{1}{A_{cor}}dR^2 \nonumber\\
    {} & \quad + \frac{u_T^2}{L^2}(dx_1^2+dx_2^2+dx_3^2)\, .\label{nhrf}
\end{align}

To obtain the metric of the far-horizon region with the inclusion of Schwarzian modes, we need to transform back to the original full-region metric \eqref{full} from this near-horizon metric \eqref{nhrf} via inverse coordinate transformation, and then take the limit $u\gg u_T$. The existence of the correction factor does not affect the validity of the previous coordinate transformation, and it propagates to the far-away region in this way. Therefore, the metric of the far-away region with the Schwarzian modes can be obtained as 
\begin{equation}
    ds^2_{FARf}=-\frac{ u^2}{L^2}\frac{1}{A_{cor}}dt^2 + \frac{L^2}{ u^2}\frac{1}{A_{cor}}du^2 + \frac{u^2}{L^2}(dx_1^2+dx_2^2+dx_3^2)\, .
\end{equation}
Compared with the classical metric \eqref{eq:FAR class metric} of the far-away region, this quantum-corrected metric acquires a correction factor $1/A_{cor}$. In the $y$ coordinate, the quantum-corrected metric of the far-away region becomes
\begin{equation}
    ds^2_{RSf}=\frac{1}{A_{cor}}dy^2+e^{-\frac{2y}{L}}(-\frac{1}{A_{cor}}dt^2+dx_1^2+dx_2^2+dx_3^2)\, ,
\end{equation}
which should be the RS metric with quantum fluctuations. Let us denote the line element and the metric as $ds^2_{RSf}$ and $G_{MN}(f)$, respectively. It is noted that $A_{cor}$ is temperature-dependent, so we have incorporated low temperatures and quantum fluctuations into the RS metric.

\section{graviton modes}\label{GM}

On the background with quantum fluctuations obtained in the previous section, we turn on perturbations in the following way:
\begin{equation}
     H_{MN}(f) = G_{MN}(f)+e^{-\frac{2y}{L}}h_{MN}\, ,
\end{equation}
subject to the constraint
\begin{equation}
     h_{M5} = 0\, ,
\end{equation}
where $h_{\mu5}$ is set to zero due to symmetry, while scalar fluctuations $h_{55}$ corresponding to the radion can be stabilized by the GW field.

The quantum-corrected equation of motion is then given by the Schwinger-Dyson equation \cite{Dyson, SCHWINGER}:
\begin{equation}
  \left\langle \frac{\delta I_m}{\delta h^{\mu\nu}}(f)\right\rangle = 0\, ,
\end{equation}
where $I_m$ is the action for $h_{\mu\nu}$. In fact, the corresponding classical equation
\begin{equation}
  \frac{\delta I_m}{\delta h^{\mu\nu}} = 0
\end{equation}
is just the linearized Einstein equation \eqref{Ein}. Hence, the quantum-corrected equation of motion has the explicit expression:
    \begin{align}
        (\nabla^2-\langle A_{cor}\rangle\partial_t\partial_t)h_{\mu\nu} + \frac{1+\langle A_{cor}\rangle}{2}e^{-\frac{2y}{L}}\partial_y\partial_yh_{\mu\nu} & {}\nonumber\\
         - \frac{1}{2}\left(3\frac{d\langle A_{cor}\rangle}{dy} + \frac{4}{L} (1+\langle A_{cor}\rangle)
        \right) e^{-\frac{2y}{L}}\partial_yh_{\mu\nu} & = 0\, .\label{eom}
    \end{align}
In principle, we need to evaluate the path integral on \eqref{A} to determine $\langle A_{cor}\rangle$. Noting $\epsilon(x^+)\epsilon(x^-)=\frac{1}{2}(\epsilon(x^+)\epsilon(x^-)+\epsilon(x^-)\epsilon(x^+))$, and considering that the one-point function of the Schwarzian mode vanishes, while the two-point function is given by \cite{Sachdev-Ye-Kitaev, Zhenbin}
\begin{align}
  {} & \langle \epsilon(x^+) \epsilon(x^-) \rangle \nonumber\\
  =\, & \frac{\beta^3}{16\pi^4 C} \Bigg[1 + \frac{\pi^2}{6} + \frac{5\cos\frac{2\pi (x^+-x^-)}{\beta}}{2} \nonumber\\
  {} & \qquad\quad - \frac{1}{2} \left(\frac{2\pi(x^+-x^-)}{\beta}-\pi\right)^2 \nonumber\\
  {} & \qquad\quad + \left(\frac{2\pi(x^+-x^-)}{\beta}-\pi\right)\sin\frac{2\pi(x^+-x^-)}{\beta}\Bigg]\, ,
\end{align}
we obtain 
\begin{align}
    {} & \langle A_{cor}\rangle \nonumber\\
    =\, & 1-\frac{\beta}{16C\pi^2} \Bigg(8 + 8\frac{\pi}{\beta}(x^+-x^-)\frac{\cos\frac{\pi}{\beta}(x^+-x^-)}{\sin\frac{\pi}{\beta}(x^+-x^-)}\nonumber\\
    {} & \qquad\quad -\frac{4\pi^2}{\beta^2}(x^+-x^-)^2\frac{\cos\frac{2\pi}{\beta}(x^+-x^-)}{\sin^2\frac{\pi}{\beta}(x^+-x^-)}\Bigg)\, .
\end{align}
Its form in the $u$ coordinate is 
\begin{align}
    {} & \langle A_{cor}\rangle \nonumber\\
    =\, & 1 - \frac{\beta}{16C\pi^2} \Bigg(8 + \frac{4(u-u_T+\delta)}{\delta} \ln\frac{u-u_T+2\delta}{u-u_T} \nonumber\\
    {} & - 3\ln^2\frac{u-u_T+2\delta}{u-u_T}{}+\frac{(u-u_T+\delta)^2}{\delta^2}\ln^2\frac{u-u_T+2\delta}{u-u_T}\Bigg)\, .
\end{align}
Taking $u\gg u_T$, we obtain the leading-order quantum correction factor in the far-away region as
\begin{equation}
    \langle A_{cor}\rangle = 1-\frac{5\beta}{4C\pi^2}+\frac{L^2}{6C\pi^2u_0}e^{\frac{y}{L}}\, .
\end{equation}
Here, only the corrections of the first two orders are retained. To solve \eqref{eom}, we apply the separation of variables:
\begin{equation}
 h_{\mu\nu}=\Sigma_{n\omega}h^n_{\mu\nu}(x)f_n^{\omega}(y)e^{-i\omega t}\, .
\end{equation}
Consequently, we obtain
    \begin{equation}
      \begin{aligned}
          \nabla^2h^n_{\mu\nu} & =-k^2_n h^n_{\mu\nu} \, ,\\
          (1+\langle A_{cor}\rangle)\frac{d^2f_n^{\omega}}{dy^2} & - \left(\frac{4}{L}(1+\langle A_{cor}\rangle)+3\frac{d\langle A_{cor}\rangle}{dy}\right)\frac{df^{\omega}_n}{dy} \\
              {} & = 2(k_n^2-\omega^2\langle A_{cor}\rangle)e^{\frac{2y}{L}}f_n^{\omega}\, .
      \end{aligned}
  \end{equation}
The boundary conditions are similar to the ones in \eqref{b}.

The mass of the KK mode $M_n$ equals the angular frequency $\omega$ when the momentum $k_n$ is zero. For $M_0=0$, there is no correction. For $M_n\neq 0$, if we only take into account the lowest order of the correction factor, then 
\begin{equation}
    M_{n}=\sqrt{\frac{8C\pi^2-5\beta}{8C\pi^2-10\beta}}\, m_n\, ,
\end{equation}
where $m_n$ is the original mass in \eqref{m}. It can be seen that the masses of the modified KK modes are larger than the original mass, and the correction is the same temperature-dependent overall factor for different values of $n$.

If we include higher-order terms in the correction factor, we can obtain the mass spectrum by numerical solution. In order to solve the hierarchy problem, we require $1/L$ to be at the Planck scale, and $\pi r_c/L=36.84$ \cite{Large}. In numerical calculations, we adopt the Planck scale, $10^{16}$ TeV, as the fundamental unit and set $L=1$. Perturbative calculations require $\beta/C$ to be less than 1, so we take a few values $0.1$, $0.9$, and $0.99$ for $\beta/C$. The limit $u\gg u_T$ means $\frac{L^2}{u_0\beta}e^{y/L}\ll \frac{2-Q^2}{6\pi}$, and we take $\frac{L^2}{u_0\beta}e^{\pi r_c/L}=0.1$, $0.05$, and $0.01$ accordingly. The numerical results are listed in the Tab.~\ref{tab:m1} through Tab.~\ref{tab:m6}. Since we take the Planck scale as the fundamental unit and the eigenvalues $M_n^2$ are extremely small, we multiply them by $10^{32}$ in practice. Therefore, the unit of $M_n$'s numerical results should be $10^{-16}\, L^{-1}$. Since $L^{-1}$ is on the Planck scale, the unit of numerical results in the table is TeV.

For the case with only the lowest-order corrections, the three different values of $\beta/C$ lead to the corrections of 0.32\%, 3.16\%, and 3.52\%, respectively.  After considering higher-order corrections, we find that the corrections vary for different modes. The maximum corrections in each table are -5.72\%, -1.85\%, -0.27\%, 0.59\%, 1.12\%, and 1.50\% in sequence. It can be observed that as $n$ increases, the correction ratio also increases. The negative correction ratios in the first three cases indicate that their masses decrease. Moreover, we observe that the effect of higher-order correction terms is opposite to that of lower-order correction terms. 
\begin{center}
\begin{table}[htb!]
    \centering
    \renewcommand{\arraystretch}{2}
    \begin{tabular}{|c|c|c|c|c|}
    \hline
           \diagbox[width=7em,          
      height=3.5em,
      font=\centering
      ]{$\beta/C$}{$\frac{e^{\pi r_c}}{u_0\beta}$}
      &uncorrected&0.01&0.05&0.1\\
         \hline
           uncorrected&3.832&\diagbox[dir=NW]{}{}&\diagbox[dir=NW]{}{}&\diagbox[dir=NW]{}{}\\
         \hline
           0.1&3.844&3.810&3.810&3.810\\
         \hline
           0.9&3.953&3.633&3.634&3.635\\
         \hline 
           0.99&3.967&3.613&3.614&3.615\\\hline
    \end{tabular}
    \caption{$m_1$ [TeV]}\label{tab:m1}
\end{table}
\end{center}

\begin{center}
\begin{table}[htb!]
    \centering
    \renewcommand{\arraystretch}{2}
    \begin{tabular}{|c|c|c|c|c|}
    \hline
           \diagbox[width=7em,          
      height=3.5em,
      font=\centering
      ]{$\beta/C$}{$\frac{e^{\pi r_c}}{u_0\beta}$}
      &uncorrected&0.01&0.05&0.1\\
         \hline
           uncorrected&7.0156&\diagbox[dir=NW]{}{}&\diagbox[dir=NW]{}{}&\diagbox[dir=NW]{}{}\\
         \hline
           0.1&7.038&7.0014&7.0015&7.0015\\
         \hline
           0.9&7.238&6.8966&6.8968&6.8970\\
         \hline 
           0.99&7.263&6.8858&6.8860&6.8863\\\hline
    \end{tabular}
    \caption{$m_2$ [TeV]}
\end{table}
\end{center}

\begin{center}
\begin{table}[htb!]
    \centering
    \renewcommand{\arraystretch}{2}
    \begin{tabular}{|c|c|c|c|c|}
    \hline
           \diagbox[width=7em,          
      height=3.5em,
      font=\centering
      ]{$\beta/C$}{$\frac{e^{\pi r_c}}{u_0\beta}$}
      &uncorrected&0.01&0.05&0.1\\
         \hline
           uncorrected&10.173&\diagbox[dir=NW]{}{}&\diagbox[dir=NW]{}{}&\diagbox[dir=NW]{}{}\\
         \hline
           0.1&10.206&10.168&10.168&10.168\\
         \hline
           0.9&10.496&10.146&10.146&10.145\\
         \hline 
           0.99&10.532&10.146&10.145&10.145\\\hline
    \end{tabular}
    \caption{$m_3$ [TeV]}
\end{table}
\end{center}

\begin{center}
\begin{table}[htb!]
    \centering
    \renewcommand{\arraystretch}{2}
    \begin{tabular}{|c|c|c|c|c|}
    \hline
           \diagbox[width=7em,          
      height=3.5em,
      font=\centering
      ]{$\beta/C$}{$\frac{e^{\pi r_c}}{u_0\beta}$}
      &uncorrected&0.01&0.05&0.1\\
         \hline
           uncorrected&13.324&\diagbox[dir=NW]{}{}&\diagbox[dir=NW]{}{}&\diagbox[dir=NW]{}{}\\
         \hline
           0.1&13.366&13.328&13.328&13.328\\
         \hline
           0.9&13.746&13.391&13.390&13.390\\
         \hline 
           0.99&13.793&13.402&13.401&13.400\\\hline
    \end{tabular}
    \caption{$m_4$ [TeV]}
\end{table}
\end{center}

\begin{center}
\begin{table}[htb!]
    \centering
    \renewcommand{\arraystretch}{2}
    \begin{tabular}{|c|c|c|c|c|}
    \hline
           \diagbox[width=7em,          
      height=3.5em,
      font=\centering
      ]{$\beta/C$}{$\frac{e^{\pi r_c}}{u_0\beta}$}
      &uncorrected&0.01&0.05&0.1\\
         \hline
           uncorrected&16.471&\diagbox[dir=NW]{}{}&\diagbox[dir=NW]{}{}&\diagbox[dir=NW]{}{}\\
         \hline
           0.1&16.523&16.485&16.485&16.485\\
         \hline
           0.9&16.992&16.635&16.634&16.632\\
         \hline 
           0.99&17.051&16.656&16.655&16.653\\\hline
    \end{tabular}
    \caption{$m_5$ [TeV]}
\end{table}
\end{center}

\begin{center}
\begin{table}[htb!]
    \centering
    \renewcommand{\arraystretch}{2}
    \begin{tabular}{|c|c|c|c|c|}
    \hline
           \diagbox[width=7em,          
      height=3.5em,
      font=\centering
      ]{$\beta/C$}{$\frac{e^{\pi r_c}}{u_0\beta}$}
      &uncorrected&0.01&0.05&0.1\\
         \hline
           uncorrected&19.616&\diagbox[dir=NW]{}{}&\diagbox[dir=NW]{}{}&\diagbox[dir=NW]{}{}\\
         \hline
           0.1&19.679&19.640&19.640&19.640\\
         \hline
           0.9&20.237&19.878&19.876&19.874\\
         \hline 
           0.99&20.307&19.910&19.908&19.906\\\hline
    \end{tabular}
    \caption{$m_6$ [TeV]}\label{tab:m6}
\end{table}
\end{center}

\section{Goldberger-Wise mechanism}\label{GW}

Now, we discuss the impact of quantum corrections on the Goldberger-Wise (GW) mechanism. First, introduce a scalar field $\Phi$, i.e., the GW field, to the model with the following bulk action 
\begin{equation}
    I_\Phi = \frac{1}{2}\int d^4x\int^{\pi r_c}_{0}dy \sqrt{G}(G^{AB}\partial_A\Phi\partial_B\Phi-M_\Phi^2\Phi^2)\, .
\end{equation}
The classical scalar field $\Phi$ satisfies the Klein-Gordon equation $\frac{\delta I_\Phi}{\delta\Phi}=0$. Under quantum corrections, according to the Schwinger-Dyson equation, we have
\begin{equation}
    \left\langle\frac{\delta I_\Phi}{\delta\Phi}\right\rangle = \langle A_{cor}\rangle e^{\frac{4y}{L}}\partial_y(e^{-\frac{4y}{L}}\partial_y\Phi)+M_\Phi^2\Phi = 0\, .
\end{equation}
We can define an effective mass 
\begin{equation}
    m_\Phi^2=\frac{M_\Phi^2}{\langle A_{cor}\rangle}\, .
\end{equation}
The GW field $\Phi$ after quantum corrections still satisfies the Klein-Gordon equation but with a renormalized mass. The boundary conditions for the equation are
\begin{equation}
    \begin{aligned}
        \Phi(y=0)&=\Phi_{UV}\, ,\\
        \Phi(y=\pi r_c)&=\Phi_{IR}\, .
    \end{aligned}
\end{equation}
If we only consider the lowest order of the quantum correction factor, the effective mass does not vary with $y$, and we can use the well-known result to obtain \cite{ WOS:000084152000004}
\begin{equation}
  \pi r_c= \frac{4}{ Lm_\Phi^2}\ln{\frac{\Phi_{UV}}{\Phi_{IR}}}=\frac{4C\pi^2-5\beta}{C\pi^2LM_\Phi^2}\ln{\frac{\Phi_{UV}}{\Phi_{IR}}}\, .
\end{equation}
With $\ln{\frac{\Phi_{UV}}{\Phi_{IR}}}$ of the order unity and $\beta/C<1$, no extreme fine-tuning of parameters is required to get the right magnitude for $\pi r_c$. For instance, taking $\frac{\Phi_{UV}}{\Phi_{IR}}=1.5$, $\frac{\beta}{C}=0.99$, and $LM_\Phi=0.196$ yields $\frac{\pi r_c}{L}=36.84$. Therefore, the GW mechanism remains valid in the presence of quantum corrections.

\section{conclusion}\label{con}

In this paper, we have considered the influence of quantum fluctuations near the horizon of a near-extremal RN black brane on the RS model. More explicitly, we have derived the RS metric by incorporating quantum fluctuations, calculated the corrections to the Kaluza-Klein mass spectrum, and discussed the impact of quantum corrections on the GW mechanism. Our work has made initial attempts to incorporate the effects of infrared quantum gravity into the RS model and, at the same time, to introduce temperature-dependent quantum effects.

It is noted that a recent work \cite{WOS:000970026700004} has also discussed the relationship between the RS model and the JT gravity, but there are differences between this work and our paper. In our paper, the JT gravity describes only the near-horizon geometry of black branes, not the braneworld model. We use JT gravity only as a tool to account for quantum fluctuations near the black brane horizons.

There are still many related issues worthy of further exploration. What corrections will this infrared quantum-gravity effect introduce into more braneworld models? What impact will the RS model with temperature have on the studies of cosmology and phase transitions? These questions await further research in the future.

\vspace{10mm}
\section*{ACKNOWLEDGMENTS}

 We would like to thank Zheng Jiang, Yu-Xiao Liu, and Lisa Randall for many helpful discussions. This work is partially supported by the International Partnership Program of the Chinese Academy of Sciences under grant No. 025GJHZ2023106GC. J.~N. was supported in part by the NSFC under grant No.~12375067, No.~12147103, and No.~12247103.

\bibliography{references}

@article{Large,
Author = {Randall, L and Sundrum, R},
Title = "{Large mass hierarchy from a small extra dimension}",
Journal = {PHYSICAL REVIEW LETTERS},
Year = {1999},
Volume = {83},
Number = {17},
Pages = {3370-3373},
Month = {OCT 25},
DOI = {10.1103/PhysRevLett.83.3370},
ISSN = {0031-9007},
ORCID-Numbers = {Sundrum, Raman/0009-0004-7537-5357},
Unique-ID = {WOS:000083242800007},
}

@article{alternative,
Author = {Randall, L and Sundrum, R},
Title = "{An alternative to compactification}",
Journal = {PHYSICAL REVIEW LETTERS},
Year = {1999},
Volume = {83},
Number = {23},
Pages = {4690-4693},
Month = {DEC 6},
DOI = {10.1103/PhysRevLett.83.4690},
ISSN = {0031-9007},
ORCID-Numbers = {Sundrum, Raman/0009-0004-7537-5357},
Unique-ID = {WOS:000084018200004},
}

@article{Zhenbin,
Author = {Maldacena, Juan and Stanford, Douglas and Yang, Zhenbin},
Title = "{Conformal symmetry and its breaking in two-dimensional nearly anti-de
   Sitter space}",
Journal = {PROGRESS OF THEORETICAL AND EXPERIMENTAL PHYSICS},
Year = {2016},
Volume = {2016},
Number = {12},
Month = {DEC},
DOI = {10.1093/ptep/ptw124},
Article-Number = {12C104},
ISSN = {2050-3911},
ResearcherID-Numbers = {yang, zhenbin/AGN-9796-2022
   },
ORCID-Numbers = {Stanford, Douglas/0000-0002-3737-794X},
}

@article{Turiaci,
Author = {Iliesiu, V, Luca and Turiaci, Gustavo J.},
Title = "{The statistical mechanics of near-extremal black holes}",
Journal = {JOURNAL OF HIGH ENERGY PHYSICS},
Year = {2021},
Number = {5},
Month = {MAY 18},
DOI = {10.1007/JHEP05(2021)145},
Article-Number = {145},
ISSN = {1126-6708},
ISSN = {1029-8479},
ORCID-Numbers = {Iliesiu, Luca/0000-0001-7567-7516
   Turiaci, Gustavo/0000-0002-1022-6287},
Unique-ID = {WOS:000762488600001},
}

@article{twoSPACETIME,
Author = {Teitelboim, C},
Title = "{Gravitation and Hamiltonian-Structure in 2 Spacetime Dimensions}",
Journal = {PHYSICS LETTERS B},
Year = {1983},
Volume = {126},
Number = {1-2},
Pages = {41-45},
DOI = {10.1016/0370-2693(83)90012-6},
ISSN = {0370-2693},
Unique-ID = {WOS:A1983QX03600011},
}

@article{JACKIW,
Author = {Jackiw, R},
Title = "{Lower Dimensional Gravity}",
Journal = {NUCLEAR PHYSICS B},
Year = {1985},
Volume = {252},
Number = {1-2},
Pages = {343-356},
DOI = {10.1016/0550-3213(85)90448-1},
ISSN = {0550-3213},
Unique-ID = {WOS:A1985AED9900029},
}

@article{Sachdev-Ye-Kitaev,
Author = {Maldacena, Juan and Stanford, Douglas},
Title = "{Remarks on the Sachdev-Ye-Kitaev model}",
Journal = {PHYSICAL REVIEW D},
Year = {2016},
Volume = {94},
Number = {10},
Month = {NOV 4},
DOI = {10.1103/PhysRevD.94.106002},
Article-Number = {106002},
ISSN = {2470-0010},
EISSN = {2470-0029},
ORCID-Numbers = {Stanford, Douglas/0000-0002-3737-794X},
Unique-ID = {WOS:000386897500004},
}

@article{Dyson,
  title = "{The $S$ Matrix in Quantum Electrodynamics}",
  author = {Dyson, F. J.},
  journal = {Phys. Rev.},
  volume = {75},
  issue = {11},
  pages = {1736--1755},
  numpages = {0},
  year = {1949},
  month = {Jun},
  publisher = {American Physical Society},
  doi = {10.1103/PhysRev.75.1736},
  url = {https://link.aps.org/doi/10.1103/PhysRev.75.1736}
}

@article{SCHWINGER,
Author = {Schwinger, J},
Title = "{On the Green's functions of quantized fields. 1.}",
Journal = {PROCEEDINGS OF THE NATIONAL ACADEMY OF SCIENCES OF THE UNITED STATES OF
   AMERICA},
Year = {1951},
Volume = {37},
Number = {7},
Pages = {452-455},
DOI = {10.1073/pnas.37.7.452},
ISSN = {0027-8424},
Unique-ID = {WOS:A1951XZ93200019},
}

@article{EXTRA1,
Author = {Rubakov, VA and Shaposhnikov, ME},
Title = "{Extra Space-Time Dimensions: Towards a Solution to the Cosmological Constant Problem}",
Journal = {PHYSICS LETTERS B},
Year = {1983},
Volume = {125},
Number = {2-3},
Pages = {139-143},
DOI = {10.1016/0370-2693(83)91254-6},
ISSN = {0370-2693},
ResearcherID-Numbers = {Shaposhnikov, Mikhail/AAF-1912-2020
   Rubakov, Valery/B-1340-2014},
ORCID-Numbers = {Shaposhnikov, Mikhail/0000-0001-7930-4565
   },
Unique-ID = {WOS:A1983QS65500012},
}

@article{DOMAIN-WALL,
Author = {Rubakov, VA and Shaposhnikov, ME},
Title = "{Do We Live Inside a Domain Wall?}",
Journal = {PHYSICS LETTERS B},
Year = {1983},
Volume = {125},
Number = {2-3},
Pages = {136-138},
DOI = {10.1016/0370-2693(83)91253-4},
ISSN = {0370-2693},
EISSN = {1873-2445},
ResearcherID-Numbers = {Shaposhnikov, Mikhail/AAF-1912-2020
   Rubakov, Valery/B-1340-2014},
ORCID-Numbers = {Shaposhnikov, Mikhail/0000-0001-7930-4565
   },
Unique-ID = {WOS:A1983QS65500011},
}

@article{WOS:000074694500008,
Author = {Arkani-Hamed, N and Dimopoulos, S and Dvali, G},
Title = "{The hierarchy problem and new dimensions at a millimeter}",
Journal = {PHYSICS LETTERS B},
Year = {1998},
Volume = {429},
Number = {3-4},
Pages = {263-272},
Month = {JUN 18},
DOI = {10.1016/S0370-2693(98)00466-3},
ISSN = {0370-2693},
Unique-ID = {WOS:000074694500008},
}

@article{WOS:000081063400007,
Author = {Maldacena, J},
Title = "{The large-N limit of superconformal field theories and supergravity}",
Journal = {INTERNATIONAL JOURNAL OF THEORETICAL PHYSICS},
Year = {1999},
Volume = {38},
Number = {4},
Pages = {1113-1133},
Month = {APR},
Note = {Quantum Gravity in the Southern Cone Conference, CTR ATOMICO BARILOCHE,
   SAN CARLOS BARILO, ARGENTINA, JAN 07-10, 1998},
DOI = {10.1023/A:1026654312961},
ISSN = {0020-7748},
ORCID-Numbers = {Fearn, Sam/0000-0003-1886-7879},
Unique-ID = {WOS:000081063400007},
}

@article{WOS:000084152000004,
Author = {Goldberger, WD and Wise, MB},
Title = "{Modulus stabilization with bulk fields}",
Journal = {PHYSICAL REVIEW LETTERS},
Year = {1999},
Volume = {83},
Number = {24},
Pages = {4922-4925},
Month = {DEC 13},
DOI = {10.1103/PhysRevLett.83.4922},
ISSN = {0031-9007},
ORCID-Numbers = {Wise, Mark/0000-0002-9125-801X},
Unique-ID = {WOS:000084152000004},
}

@article{WOS:001251818200003,
Author = {Mishra, Rashmish K. and Randall, Lisa},
Title = "{Phase transition to RS: cool, not supercool}",
Journal = {JOURNAL OF HIGH ENERGY PHYSICS},
Year = {2024},
Number = {6},
Month = {JUN 18},
DOI = {10.1007/JHEP06(2024)099},
Article-Number = {99},
ISSN = {1029-8479},
ResearcherID-Numbers = {Mishra, Rashmish/IAP-6051-2023},
Unique-ID = {WOS:001251818200003},
}

@article{WOS:001116716600002,
Author = {Mishra, Rashmish K. and Randall, Lisa},
Title = "{Consequences of a stabilizing field's self-interactions for RS cosmology}",
Journal = {JOURNAL OF HIGH ENERGY PHYSICS},
Year = {2023},
Number = {12},
Month = {DEC 5},
DOI = {10.1007/JHEP12(2023)036},
Article-Number = {36},
ISSN = {1029-8479},
ResearcherID-Numbers = {Mishra, Rashmish/IAP-6051-2023},
Unique-ID = {WOS:001116716600002},
}

@article{WOS:000423800200001,
Author = {Bunk, Don and Hubisz, Jay and Jain, Bithika},
Title = "{A perturbative RS I cosmological phase transition}",
Journal = {EUROPEAN PHYSICAL JOURNAL C},
Year = {2018},
Volume = {78},
Number = {1},
Month = {JAN 29},
DOI = {10.1140/epjc/s10052-018-5529-2},
Article-Number = {78},
ISSN = {1434-6044},
EISSN = {1434-6052},
ORCID-Numbers = {Hubisz, Jay/0000-0002-5944-5413},
Unique-ID = {WOS:000423800200001},
}

@article{WOS:000175277100051,
Author = {Creminelli, P and Nicolis, A and Rattazzi, R},
Title = "{Holography and the electroweak phase transition}",
Journal = {JOURNAL OF HIGH ENERGY PHYSICS},
Year = {2002},
Number = {3},
Month = {MAR},
DOI = {10.1088/1126-6708/2002/03/051},
Article-Number = {051},
EISSN = {1029-8479},
ORCID-Numbers = {Creminelli, Paolo/0000-0003-0491-3328
   Rattazzi, Riccardo/0000-0003-0276-017X
   Nicolis, Alberto/0000-0003-2024-6203},
Unique-ID = {WOS:000175277100051},
}

@misc{liu2024quantumcorrectionsholographicstrange,
      title="{Quantum Corrections to Holographic Strange Metal at Low Temperature}",
      author={Xiao-Long Liu and Jun Nian and Leopoldo A. Pando Zayas},
      year={2024},
      eprint={2410.11487},
      archivePrefix={arXiv},
      primaryClass={hep-th},
      url={https://arxiv.org/abs/2410.11487}, 
}

@misc{liu2025quantumcorrectedholographicwilsonloop,
      title="{Quantum-Corrected Holographic Wilson Loop Expectation Values and Super-Yang-Mills Confinement}", 
      author={Xiao-Long Liu and Cong-Yuan Yue and Jun Nian and Wenni Zheng},
      year={2025},
      eprint={2412.11107},
      archivePrefix={arXiv},
      primaryClass={hep-th},
      url={https://arxiv.org/abs/2412.11107}, 
}

@misc{zheng,
      title="{Quantum Gravity Corrections to the Scalar Quasi-Normal Modes in Near-Extremal Reissener-Nordstr\"om Black Holes}",
      author={Zheng Jiang and Jun Nian and Caiying Shao and Yu Tian and Hongbao Zhang},
      year={2025},
      eprint={2506.22945},
      archivePrefix={arXiv},
      primaryClass={hep-th},
      url={https://arxiv.org/abs/2506.22945}, 
}

@misc{nian2025quantumcorrectionslowtemperaturefluidgravity,
      title="{Quantum Corrections in the Low-Temperature Fluid/Gravity Correspondence}",
      author={Jun Nian and Leopoldo A. Pando Zayas and Cong-Yuan Yue},
      year={2025},
      eprint={2510.15411},
      archivePrefix={arXiv},
      primaryClass={hep-th},
      url={https://arxiv.org/abs/2510.15411}, 
}

@misc{cremonini2025quantumcorrectionsetasjt,
      title="{Quantum Corrections to $\eta/s$ from JT Gravity}",
      author={Sera Cremonini and Li Li and Xiao-Long Liu and Jun Nian},
      year={2025},
      eprint={2510.21602},
      archivePrefix={arXiv},
      primaryClass={hep-th},
      url={https://arxiv.org/abs/2510.21602}, 
}

@article{WOS:000970026700004,
Author = {Geng, Hao and Karch, Andreas and Perez-Pardavila, Carlos and Raju, Suvrat and Randall, Lisa and Riojas, Marcos and Shashi, Sanjit},
Title = "{Jackiw-Teitelboim Gravity from the Karch-Randall Braneworld}",
Journal = {PHYSICAL REVIEW LETTERS},
Year = {2022},
Volume = {129},
Number = {23},
Month = {NOV 29},
DOI = {10.1103/PhysRevLett.129.231601},
Article-Number = {231601},
ISSN = {0031-9007},
EISSN = {1079-7114},
ResearcherID-Numbers = {SekharRaju, Sankara/KEI-7216-2024
   Geng, Hao/HJP-9895-2023
   Karch, Andreas/AAD-2221-2022
   Riojas, Marcos/JNT-1602-2023
   },
ORCID-Numbers = {Riojas, Marcos/0000-0002-2411-1333
   Geng, Hao/0000-0001-9251-5762
   Perez-Pardavila, Carlos/0000-0003-4986-3299
   Karch, Aneas/0000-0002-5725-2124
   Raju, Suvrat/0000-0002-0675-5661
   Shashi, Sanjit/0000-0002-4856-5890},
Unique-ID = {WOS:000970026700004},
}

@misc{Brown:2024ajk,
      title="{The evaporation of charged black holes}", 
      author={Adam R. Brown and Luca V. Iliesiu and Geoff Penington and Mykhaylo Usatyuk},
      year={2024},
      eprint={2411.03447},
      archivePrefix={arXiv},
      primaryClass={hep-th},
      url={https://arxiv.org/abs/2411.03447}, 
}

@article{Emparan:2025sao,
    author = "Emparan, Roberto",
    title = "{Quantum cross-section of near-extremal black holes}",
    eprint = "2501.17470",
    archivePrefix = "arXiv",
    primaryClass = "hep-th",
    doi = "10.1007/JHEP04(2025)122",
    journal = "JHEP",
    volume = "04",
    pages = "122",
    year = "2025"
}

@misc{PandoZayas:2025snm,
      title="{One-loop Corrected Holographic Shear Viscosity to Entropy Density Ratio at Low Temperatures}", 
      author={Leopoldo A. Pando Zayas and Jingchao Zhang},
      year={2025},
      eprint={2510.16100},
      archivePrefix={arXiv},
      primaryClass={hep-th},
      url={https://arxiv.org/abs/2510.16100}, 
}
\end{document}